\ifdefined\drafting
\documentclass[prb,nofootinbib,onecolumn,floatfix,superscriptaddress]{revtex4-2}
\else
\documentclass[prb,nofootinbib,twocolumn,floatfix,superscriptaddress]{revtex4-2}
\fi

\usepackage{braket}
\usepackage{amsmath}
\usepackage{siunitx}
\usepackage{graphicx}
\usepackage[colorlinks=true,linkcolor=blue]{hyperref}
\usepackage[capitalise]{cleveref}
\usepackage{float}
\usepackage{bm}
\usepackage[dvipsnames,svgnames,x11names]{xcolor}

\DeclareMathOperator{\im}{Im}
\DeclareMathOperator{\re}{Re}

\ifdefined\drafting
	\usepackage[a4paper,left=0.25in,right=1.8in,head=21.0pt]{geometry}
\else
\fi

\ifdefined\showcomments{}
	\usepackage[commentmarkup=todo,todonotes={textsize=scriptsize,textwidth=1.6in}]{changes}
	\makeatletter \def\@captype{figure} \makeatother 
\else
	\usepackage[final]{changes}
\fi
\newcommand{\markhigh}[1]{\bgroup\markoverwith				
  {\textcolor{#1}{\rule[-.5ex]{2pt}{2.5ex}}}\ULon}

\definechangesauthor[name=Elina, color=ForestGreen]{EP}
\definechangesauthor[name=Slava, color=blue]{VK}
\definechangesauthor[name=Patrik, color=red]{PR}
\definechangesauthor[name=Peter, color=orange]{PS}
\definechangesauthor[name=Girts, color=cyan]{GB}

\begin{document}
\title{Collision of two interacting electrons on a mesoscopic beamsplitter: exact solution in the classical limit}
\author{Elina Pavlovska}
\affiliation{Department of Physics, University of Latvia, Riga, LV-1004, Latvia}
\author{Peter G. Silvestrov}
\affiliation{Institut f\"ur Mathematische Physik, Technische Universit\"at Braunschweig, D-38106 Braunschweig, Germany}
\author{Patrik Recher}
\affiliation{Institut f\"ur Mathematische Physik, Technische Universit\"at Braunschweig, D-38106 Braunschweig, Germany}
\affiliation{Laboratory for Emerging Nanometrology Braunschweig, D-38106 Braunschweig, Germany}
\author{Girts Barinovs}
\affiliation{Department of Physics, University of Latvia, Riga, LV-1004, Latvia}
\author{Vyacheslavs Kashcheyevs}
\affiliation{Department of Physics, University of Latvia, Riga, LV-1004, Latvia}
\begin{abstract}
   
 Experiments on collisions of isolated electrons guided along the edges in quantum Hall setups 
can mimic mixing of photons with the important distinction that electrons are charged fermions. In the so-called electronic Hong-Ou-Mandel (HOM) setup uncorrelated pairs of electrons are injected towards a beamsplitter. If the two electron wave packets were identical, Fermi statistics would force the electrons to scatter to different detectors, yet this quantum antibunching may be confounded by Coulomb repulsion. Here we model an electronic HOM experiment using a quadratic 2D saddle point potential for the beamsplitter and unscreened Coulomb interaction between the two injected electrons subjected to a strong out-of-plane magnetic field. We show that classical equations of motion for the drift dynamics of electrons’ guiding centers take on the form of Hamilton equations for canonically conjugated variables subject to the saddle point potential and the Coulomb potential where the dynamics of the center-of-mass coordinate and the relative coordinate separate. We use these equations to determine collision outcomes in terms of a few experimentally tuneable parameters: the initial energies of the uncorrelated electrons, relative time delay of injection and the shape of the saddle point potential. A universal phase diagram of deterministic bunching and antibunching scattering outcomes is presented with a single energy scale characterizing the increase of the effective barrier height due to interaction of coincident electrons.
We suggest clear-cut experimental strategies to detect the predicted effects and give analytical estimates of conditions when the classical dynamics is expected to dominate over quantum effects.
\end{abstract}
\maketitle

\section{Introduction}
Solid-state electron quantum optics
is a branch of quantum technologies and concerns the creation, characterization and exploitation of individual excitations of electrical current.  It offers potential applications in sensing, metrology and quantum information processing~\cite{Bocquillon2014,Bauerle2018,Kataoka2021,Laucht2021,Edlbauer2022}.
In direct analogy with photonics~\cite{HOM1987}, 
a hallmark signature of quantum statistics in electron quantum optics is the electronic Hong-Ou-Mandel (HOM) two-particle interference at a beamsplitter, first demonstrated~\cite{Bocquillon2013,Dubois2013} for on-demand sources of well-screened excitations of chiral edge states in an integer filling factor quantum Hall system~\cite{Feve2007}. Yet an essential difference of electrons from photons is not only the fermionic nature of the former, but also the possibility of strong Coulomb interaction if electrons are confined or propagating in isolation, as is the case for tuneable-barrier quantum dot sources~\cite{Kaestner2015,Bauerle2018,Kataoka2021} injecting electrons on demand into depleted ballistic nanostructures~\cite{Fletcher2012,Ubbelohde2015,Fletcher2019,Takada2019,Freise2019}.  Distinguishing quantum correlations due to indistinguishability of non-interacting particles from correlations caused by interactions is an essential conundrum of nanoscale quantum transport in general~\cite{Blanter1999}, and remains an open challenge for the ballistic few-electron devices in particular.
Even though the experiment of Ubbelohde \emph{et al.}~\cite{Ubbelohde2015} has reported a tantalizing bunching anomaly in partitioning of electron pairs emitted on demand, and attributed this anomaly to interactions on the beamsplitter using very general arguments, a systematic understanding of interplay between partitioning and interactions in two-electron collisions is lacking. In this paper we address part of this problem theoretically by considering a limit of strong interactions that is complementary to a much-better understood problem of HOM interferometry with non-interacting electrons~\cite{Bocquillon2014,Roussel2017}. We consider the regime of long-range two-body interactions relevant for isolated electrons in depleted edge channels which is different from many-body   physics leading to fracitonalization of excitations in HOM experiments with   quantum Hall edge channels modelled as interacting 1D quantum liquids~\cite{Wahl2013,Freulon2015,Marguerite2016decoherence,Ferraro2018,Rebora2020}.

Available analytic approaches to quantum scattering of two interacting particles on a local structure  either exploit exactly solvable limits of point-like interactions in 1D 
\cite{Aharony2e1999,Entin00EPL,Dhar2008} or are perturbative in the interaction strength \cite{Goorden2007,RyuSim2022}.
Bellentani \emph{et al.}~\cite{Bellentani2019} have explored numerically collision of two electrons at a two-dimensional (2D) constriction, looking for interaction-induced changes in the anti-bunching probabilities. A theoretical study of  single-electron emission by a time-dependent smooth potential by Ryu~\emph{et.al}~\cite{Ryu2016} has demonstrated, in particular, feasibility of a classical approximation for the motion of the guiding centre of a Gaussian wave-packet localized by strong magnetic fields. A common challenge for numerical modelling and  physical experiments is the large dimensionality of the parameter space which is difficult to explore systematically.

Recently, we have developed \cite{Silvestrov2022} a theory of two-electron effects in
electron quantum optics setups in the
strong coupling limit where the Coulomb repulsion is strong enough
to change the trajectories of two electrons. In the
present work we apply this approach for an in-depth analysis of 
the classical two-electron correlations in a HOM setup with the two
electrons colliding at a constriction which serves as a
beamsplitter (energy-filtering barrier). The constriction is modelled as a 2D quadratic saddle-point potential~ \cite{Buttiker1990} in a magnetic field perpendicular to the plane \cite{Fertig1987}. 
We treat the particles classically on the scale larger than the quantum uncertainty and wave-function overlap, and compute a  universal phase diagram of deterministic scattering outcomes as function of the incoming electrons' energies,  relative time delay, and the three parameters of the constriction (dispersion timescale $\omega^{-1}$, maximal interaction energy $U$ and the aspect ratio of the saddle). We derive experimentally testable scaling relations, and illustrate possible qualitative signatures of the interactions-dominated regime in experimentally relevant coordinates. Finally, we show how to estimate feasibility of reaching the relevant regime of $U/(\hbar \omega) \gg 1$ using the microscopic parameters of the constriction potential, magnetic confinement and the Coulomb law constant. 

The results of this study will hopefully help to map out future theoretical and experimental explorations of few-electron solid state quantum technologies with on-demand isolated wave-packets in the strong coupling regime. 

The paper is structured as follows. We start in the Section~\ref{sec:model} with the definition of the problem, the Hamiltonian  and classical equations of motion, then solution of the problem is developed in Section~\ref{sec:solution}. A reader interested primarily in the physical interpretation of the scattering solution may proceed from Section \ref{sec:schematic} directly to Section~\ref{sec:mainresults}  where the phase diagram and the potential experimental signatures are discussed.
In Section~\ref{sec:applicability} we discuss applicability of the classical approximation and conditions for neglecting quantum uncertainties and statistics. 
Finally, in Section~\ref{sec:conclusions} we put the results into a broader context of current experimental developments and sketch an outlook. Extensive Appendices at the end of the paper provide theoretical justifications for the choice of approximations and limits of applicability.

\begin{figure*}
    \centering
   \includegraphics[width=0.75\textwidth]{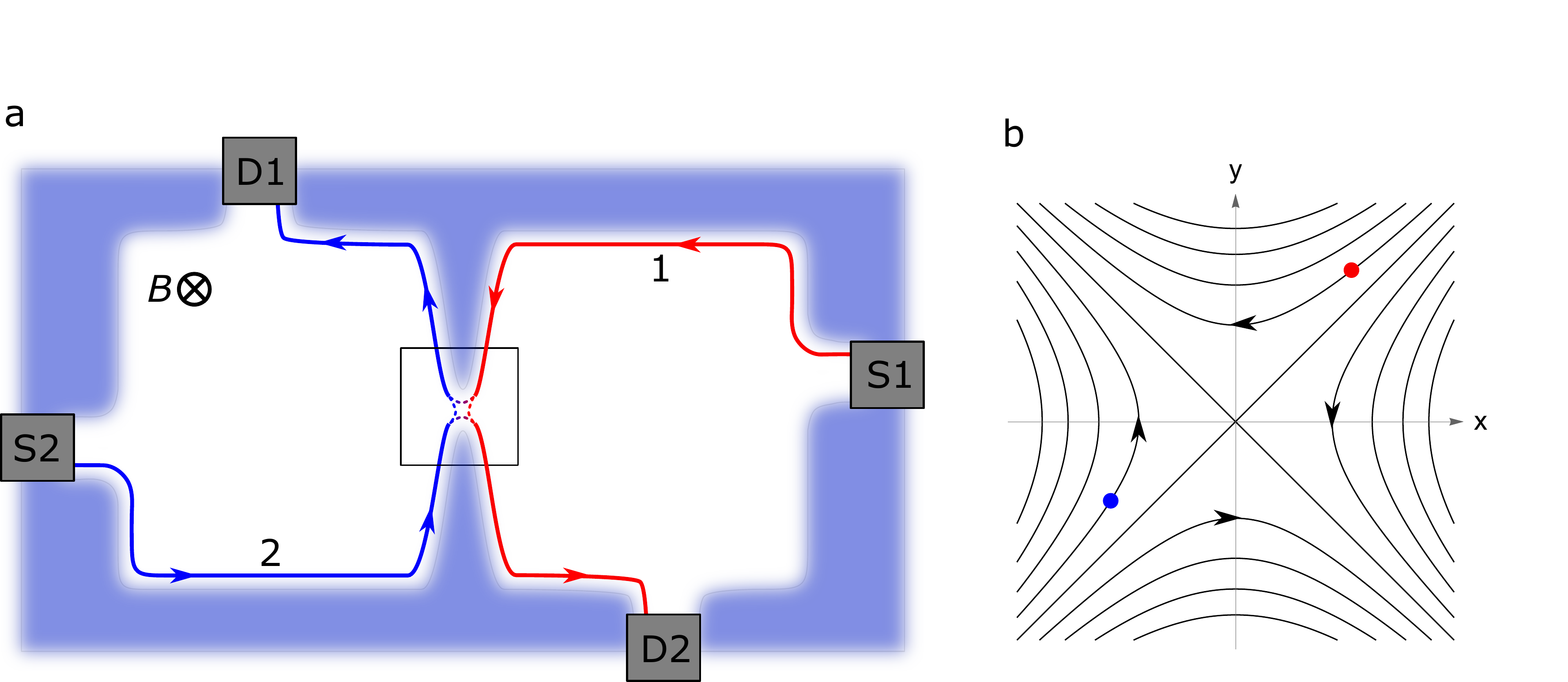}
    \caption{(a) Sketch of an experimental setup for investigation of two-electron collisions. (b) Coordinate axes and the level lines of the saddle point potential. Two colored dots  indicate the incoming electrons, and the arrows on the level lines show the direction of drift motion in the absence of interactions.}
    \label{fig:setup}
\end{figure*}

\section{Model\label{sec:model}}

\subsection{Schematic setup for a collision experiment\label{sec:schematic}}

A conceptual sketch of the experimental setup is shown in Fig.~\ref{fig:setup}a.
The sample is a 2D quantum Hall system in a strong perpendicular magnetic field. The 2D bulk electron gas (two large white areas) is depleted away from the edges (where on-demand hot electrons propagate chirally as indicated by arrows) and remains disconnected from sources and detectors at all times, make a two-body approach feasible. 
Two sources, S1 and S2, launch electrons on demand \cite{Leicht2011,Fletcher2012} at well-defined energies $\epsilon_1$ and $\epsilon_2$ with a controlled time-delay $\Delta t$ (up to unavoidable quantum uncertainty, see Refs.~\cite{Ryu2016,Kashcheyevs2017,Fletcher2019} and Section~\ref{sec:applicability} below). The electrons scatter on a central constriction (marked by a square box in Fig.~\ref{fig:setup}a) and then reach either of the two detectors, $D1$ and $D2$. The number of electrons detected at each detector is the scattering outcome. The total number of electrons in a single-shot realization is two, hence there are only three experimentally distinguishable outcomes: either $0$, $1$ or $2$ electrons at $D1$. Conventionally~\cite{Blanter1999,Kaestner2015}, the three outcomes are distinguished by repeating the experiment at a suitably chosen frequency (typically, tens to hundreds of megahertz) and measuring the zero-frequency current \cite{Fletcher2019} and crosscorrelation noise~\cite{Ubbelohde2015}. Recent advances in compatible single-shot electron counting detectors~\cite{Freise2019} would enable direct realization of our idealized $D1$ and $D2$.

\subsection{Hamiltonian}
We consider a partitioning barrier for isolated on-demand electrons described by a saddle potential in 2D,
\begin{align} \label{eq:V_saddle}
   V_{\text{saddle}}(x,y) = \frac{m}{2} \left (
    \omega_y^2 \, y^2_{} -\omega_x^2 \, x^2
    \right ) \, .
\end{align}
Level lines of $V_{\text{saddle}}(x,y)$ are shown schematically in Fig.~\ref{fig:setup}b. 

The Hamiltonian of the $j$-th electron ($j=1,2$) is 
\begin{align} \label{eq:Hsp}
  \mathcal{H}_j= \frac{1}{2 m}
    \left (\bm{p}_j + e \mathbf{A}_j\right )^2 + V_{\text{saddle}}(x_j,y_j) \, ,
\end{align}
where $\bm{p}_j = - i \hbar \{ \partial_{x_j} , \partial_{y_j}  \}$ is the canonical momentum, and the vector potential
$\mathbf{A}_j=B \left \{+y_j/2, -x_j/2 \right\}$ describes uniform magnetic field with induction $B=m \omega_c /e>0$, directed along the negative $z$-axis. Here $e$ is the elementary charge, $m$ is the effective mass and $\omega_c$ is the cyclotron frequency.
Single-particle scattering problem  for $\mathcal{H}_j$ admits an exact solution~\cite{Fertig1987} for arbitrary $\omega_x$, $\omega_y$ and $\omega_c$, see Appendix \ref{sec:FH}.

Two-electron interaction is described by 
the total Hamiltonian $\mathcal{H}=\mathcal{H}_1+\mathcal{H}_2 + V_{\text{int}}(r) $ with 
 a central two-body potential $V_{\text{int}}$ that is a function of the relative distance $r=\sqrt{(x_1-x_2)^2+(y_1-y_2)^2}$ only. We focus on a long-range Coulomb potential, 
\begin{align} \label{eq:CoulonbPure}
    V_{\text{int}}(r) = \frac{e^2}{4 \pi \varepsilon_0 \varepsilon r}  \, .
\end{align}


The Coulomb potential 
 can be parametrised as $ V_{\text{int}}(r)= {U \, d_0}/{r} $  where $U=V_{\text{int}}(d_0)=V_{\text{saddle}}(0, d_0)$ and 
\begin{align}
    d_0 =[e^2/(2 \pi \varepsilon_0 \varepsilon \, m \, \omega_y^2) ]^{1/3} 
\end{align}
 is the distance of zero net electric force acting in the transverse ($y$) direction on each electron as they pass each other along $x$:   $\partial_{y_j} [ V_{\text{saddle}}+V_{\text{int}}]=0$ for $x_1=x_2$ yields $|y_2-y_1|=d_0$. The rationale for this parametrization comes from the form of the equations of motion discussed below. We will show that $d_0$ is the minimal distance and $U$ is the maximal interaction energy in a two-electron collision in the classical limit.

\subsection{Equations of motion, conserved quantities, and dimensional crossover}

    In the large magnetic field limit, $\omega_x, \omega_y \ll \omega_c$, the electric potentials do not cause transition between Landau levels, and the classical motion of the guiding centre is chiral, described by first order differential equations of $\bm{E}\times \bm{B}$ drift along the equipotential lines\footnote{Identification of the microscopic and the guiding centre coordinates is justified within one Landau level as we explicitly demonstrate in Appendix~\ref{sec:FH}.},
\begin{align} \label{eq:vDrfit}
    \{ \dot{x}_j , \dot{y}_j \} =  \frac{\{ 
    -\partial/\partial y_j, +\partial/\partial x_j \}}{m \omega_c} \left [
    V_\text{saddle}(x_j,y_j) + V_{\text{int}}(r) \right] \, .
\end{align}

In terms of relative and center-of-mass coordinates, 
$\{ x,y\}=\{x_2-x_1,y_2-y_1\}$ and
$\{ x_{\text{c.m.}}, y_{\text{c.m.}} \} =\{ (x_1+x_2)/2,  (y_1+y_2)/2 \}$,
 equations of motion \eqref{eq:vDrfit} for the quadratic potential \eqref{eq:V_saddle} separate~\cite{Silvestrov2022}:
\begin{subequations} \label{eq:EOMseparated}
\begin{align}
  \dot{x}_{\text{c.m.}}  &= - \omega \, y_{\text{c.m.}} / \kappa  \, , \\
  \dot{y}_{\text{c.m.}}  &= - \omega \, \kappa \, x_{\text{c.m.}}\, , \\
   \dot{x} &=  \frac{\omega}{\kappa} \, y \Big( -1+\frac{d_0^{3}}{(x^2+y^2)^{3/2}} \Big) \, , \label{eq:xdot} \\
  \dot{y} &= \omega \kappa \, x \Big( -1-\frac{d_0^{3}}{\kappa^2 (x^2+y^2)^{3/2}} \Big) \, , \label{eq:ydot}
\end{align}
\end{subequations}
where $\kappa=\omega_x/ \omega_y$  and $\omega=\omega_x \omega_y /\omega_c $.
We see that the drift motion is completely specified by two dimensionful and one dimensionless parameters: the beamsplitter timescale $\omega^{-1}$, the interaction lengthscale $d_0$, and a geometric aspect ratio of the saddle $\kappa$.

Equations \eqref{eq:vDrfit} can be seen as Hamilton equations of two one-dimensional degrees of freedom with ($x_j$, $y_j$)  being the conjugate coordinate-momentum pairs in appropriate units. The corresponding quantum commutator (and hence the short-distance cut-off for classical dynamics) $[ x_j, y_j] = i l_c^2= i \hbar/( m \omega_c)$ is set by the magnetic length $l_c$ (see   Appendix \ref{sec:derive1D}).
The classical Hamiltonians leading to the separated equations of motion  \eqref{eq:EOMseparated} are also the conserved quantities,
\begin{align}
    E_{\text{c.m.}} & = m \omega_y^2 (\, y_{\text{c.m.}}^2 - \kappa^2 \, x_{\text{c.m.}}^2) \, , \label{eq:cons1} \\
    E_{+} & =m \omega_y^2  ( \, y^2 - \kappa^2 \, x^2)/4 + U \, d_0/( \sqrt{x^2+y^2}) \, . \label{eq:cons2}
\end{align} 
We have chosen constant prefactors 
in Eqs.~\eqref{eq:cons1}--\eqref{eq:cons2} to match the normal energy units; $E_{\text{c.m.}}+E_{+}$ is the total potential energy, yet the two quantities are conserved separately due to separation of variables. While $E_{+}$ is simply the energy associated with the relative coordinate (note the factor $m/4$ instead of $m/2$  due to reduced mass $\mu=m/2$), we use the subscript `$+$', since together with  a similarly defined (yet non-conserved)  quantity $E_{-}$, see Eq.~\eqref{eq:EplusIni} below, it turns useful to express our main results in Section~\ref{sec:mainresults}.

Even though the drift velocity equations are usually 
derived in the large magnetic field limit, they can be used to examine the full crossover from magnetic $(\omega_c \gg \omega_y)$ to electrostatic $(\omega_y \gg \omega_c)$ confinement in the constriction, i.e.\ from 2D chiral to 1D linear motion.
For $\omega_y \sim \omega_c$ the transverse electric field due to the term $\propto \omega_y^2 y^2$ in $V_{\text{saddle}}$ contributes not only to the drift motion but also to the quantum confinement. Indeed, as we show in Appendix \ref{sec:map},  a more general derivation leads to the same equations \eqref{eq:EOMseparated} if $(\omega, \kappa, d_0)$ are rescaled to
\begin{align} \label{eq:rescaled}
    \omega'  =\omega \, \Xi \, , \;  
    \kappa'  =  \kappa \, \Xi \, , \; 
    d_0'  =  d_0 \, \Xi^{2/3} 
\end{align}
with $\Xi = \omega_c/ \sqrt{\omega_c^2 +\omega_y^2}$ as long as $\omega_x \ll \max(\omega_c, \omega_y)$ ensures the separation of energy scales between the drift and the confined motion. 

The limit $\Xi \approx \omega_c/\omega_y \to 0$  admits reinterpretation of Eqs.~\eqref{eq:EOMseparated} as 1D Coulomb scattering~\cite{Abramovici2009} for which the magnetic field is irrelevant. Indeed, using Eq.~\eqref{eq:rescaled} to take $\Xi \to 0$,  Eq.~\eqref{eq:ydot}  becomes simply the Newton's second law,  
$\dot{\mathrm{p}} = \mu \omega_x^2 x -
\partial_x V_{\text{int}}(x) $
if identify $-y \omega_y^2/\omega_c = \mathrm{p}/\mu$ with the linear momentum governed by Eq.~\eqref{eq:xdot}, $\mathrm{p}=\mu \dot{x}$ (here $\mu=m/2$ is the reduced mass).
In this limit $\omega'=\omega_x$, but $U'= U \Xi^{-2/3} \to \infty$ as electrons cannot pass each other classically. Hence  instead of $U$, a  measure of interaction strength that does not involve $\omega_y$ is more appropriate in the 1D limit, as we will find out in the analysis of the narrow-constriction limit,
\begin{align} \label{eq:U1D}
    U_{\text{1D}} = U \kappa^{2/3} = (m/2)^{1/3} \, [\omega_x \, e^2/(4 \pi \varepsilon_0 \varepsilon) ]^{2/3} \, .
\end{align}
We return to the discussion of the competition between 1D and 2D effects in Sec.~\ref{sec:mainresults} but for the main part of the paper we consider the implications of the classical Eqs.~\eqref{eq:EOMseparated} treating the interaction strength $U$ (or $d_0$ when discussing lengths), curvature of the saddle $\omega$, and the aspect ratio of the constriction $\kappa$ as given parameters.

\section{Classical solution of the collision problem\label{sec:solution}}

\subsection{Initial conditions for the collision problem}

Two electrons are entering the scattering region from opposite quadrants: in the far past  $(x_1>0, y_1>0)$  and $(x_2<0, y_2<0)$.  Individual  energies of incoming electrons,
\begin{align}
\epsilon_j=\lim\limits_{t\to -\infty}
V_{\text{saddle}}(x_j,y_j) \, ,
\end{align}
are well defined as asymptotically  (at $|x_j|, |y_j| \gg d_0$) interactions are negligible.

Besides $\epsilon_1$ and $\epsilon_2$, the third crucial parameter is the relative time delay $\Delta t$.
If electron $2$ (electron $1$) enters the scattering region first then the injection time delay $\Delta t>0$ ($\Delta t<0$). 
In our case of  quadratic  $V_{\text{saddle}}(x,y)$ the delay $\Delta t$
can be expressed in terms of the incoming trajectory asymptotics as
\begin{align} \label{eq:timedelay}
    e^{\omega \Delta t} = \lim_{t\to -\infty} \frac{x_1(t)}{-x_2(t)} \, .
\end{align}
This equation is derived by considering the last-to-arrive electron at a point $(x_0,y_0)$ far enough from the origin and from the other electron for interactions to be negligible, yet already sufficiently close for the saddle approximation to be applicable, and then using the non-interacting solution to extrapolate the motion into far past.

The values of the conserved integrals of motion ~\eqref{eq:cons1}-\eqref{eq:cons2} are  determined by the non-interacting incoming asymptotes:
\begin{align}
    E_{\text{c.m.}} &= \epsilon_1+\epsilon_2- E_{+} \, , \label{eq:consefvationofEnergy} 
    \\
    E_{\pm} & = \frac{\epsilon_2}{2}\left (1+e^{+\omega \Delta t}\right ) \pm \frac{\epsilon_1}{2}\left (1 +e^{-\omega \Delta t}\right ) \, .  \label{eq:EplusIni}
\end{align}
Here we have additionally defined $E_{-}$ which is not a conserved quantity but will turn out to be a useful combination of initial conditions. Notation is motivated by the fact that
for coincident arrival ($\Delta t=0$) we have simply $E_{\pm}=\epsilon_{2} \pm \epsilon_{1}$.

We note that the two conservation laws alone are not sufficient to solve the scattering problem: there is an additional constraint that involves $d_0$ and $\kappa$ in a non-trivial manner which sets the relation between energy transfer and the time delay  in the outgoing asymptotes.


\subsection{Solution for the relative coordinate\label{eq:relsol}}
We first consider evolution of the relative coordinate from an initial condition $x(0)=-x_0 <0$ with $x_0 \gg d_0$ such that the contribution of the interaction term to the equations of motion \eqref{eq:EOMseparated} can be neglected initially.  
 The other initial value $y(0)<0$ is determined by the initial value of the conserved energy of relative motion $E_{+}$. As we consider the scenario of both electrons approaching the barrier 
 and  getting closer to each other up to a distance of order $d_0$, interaction effects may become relevant for $|E_{+}| \ll E_0$ where  $E_0=m \omega_x^{2} x_0^2/4$. We consider  all energies to the first relevant order and take the limit $x_0, E_0 \to \infty$ at the end.
Under these conditions the initial value $y(0)$ is given by the linearization of  \eqref{eq:cons2} with respect to $| \epsilon_1| , |\epsilon_2| \ll E_0$,
$y(0) = -\kappa x_0 (1+E_{+}/[{2 E_0}])$.

\begin{figure}
    \centering
    \includegraphics[width=0.45\textwidth]{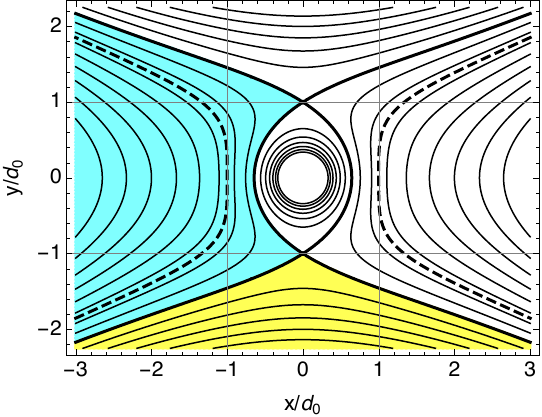}
    \caption{Relative coordinate follows the level lines of $E_{+}(x,y)$ depicted here at intervals of $0.25 \, U$ for the aspect ratio parameter $\kappa=0.5$. The thick line at $E_{+}=E_c =1.5 \, U$ separates electrons that come closest along $x$ (open region on the left, in cyan) from electrons that come closest along $y$ (open region at the bottom, in yellow). The dashed level line at $E_{+}=E_{c2}= U \, (1-\kappa^2/2)$ has a vanishing second derivative along $x$ at $\{ \pm d_0, 0 \}$. }
    \label{fig:levelsrel}
\end{figure}

Qualitative nature of trajectories near the interactions-dominated region is apparent from the level-lines plot of $E_{+}(x,y)$ \cite{Silvestrov2022} as shown in Fig.~\ref{fig:levelsrel}. The  trajectories in $\{x, y\}$ plane can be computed analytically by solving a cubic equation yielding cumbersome yet computationally efficient algebraic expressions. In the range   $E_{c2}=U (1-\kappa^2/2)  < E_{+} < E_{c} =3 U/2$ the function $x(y)$  has two minima, and $x(t)$ has one, three and zero extrema for $E_{+}<E_{c2}$, $E_{c2} < E_{+} < E_{c}$ and $E_{+}> E_{c}$, respectively. The maximal value of $x$ at $E_{+}=E_{c2}$ is equal to $-d_0$, see the dahsed line in Fig.~\ref{fig:levelsrel}.

We define the time $\tau >0$ for the relative coordinate to travel from $x(0)=-x_0$ back to a large distance $|x(\tau)|=x_0$.
There are two possibilities,
\begin{align} \label{eq:taudef}
    x(\tau)=\begin{cases} + x(0) \, , & E_{+} <E_c \, , \\
    -x(0) \, , & E_{+} > E_c \, , \end{cases}
\end{align}
that correspond to reflection or transmission of the relative coordinate, respectively (see color-shaded regions in  Fig.~\ref{fig:levelsrel}). As the velocity vector $(\dot{x}, \dot{y})$ is a
tangent to the level lines,  the  critical value $E_{+}=E_c$ corresponding to degeneracy of reflection  with transmission can be found by setting $\dot{x}=\dot{y}=0$ in the equations of motion~\eqref{eq:xdot} and \eqref{eq:ydot}. This gives $E_c= 3 \, U /2$ and $(0,-y_0)$ as the location of the critical point (approached from negative $y$), see Figure \ref{fig:levelsrel}.

At  $E_{+}=E_c$ the relative coordinate trajectory approaches the unstable equilibrium point asymptotically, along the boundary between the two shaded regions in Fig.~\ref{fig:levelsrel}. As a consequence, 
 $\tau \to \infty$ at $E_{+}=E_c$ even  for finite $E_0$ (for $E_0 \to \infty$, $\tau\to \infty$ simply because the starting coordinate $x_0$ moves infinitely far away).

In the non-interacting case ($U=0$), the travel time for large $E_0$ equals to
\begin{align} \label{eq:nonint}
    \tau_{U\!=\!0} =\omega^{-1} \ln |  4 \, E_0/E_{+} |  \, ,
\end{align} 
which 
diverges logarithmically both for large $E_0$ and for $|E_{+}| \to 0$ (which is $E_c$ for $U=0$). 

For finite $U$, we compute $\tau$ via numerical quadratures and express the results in terms of
a dimensionless function $\Phi_{\kappa}(z)$,
\begin{align}
    \Phi_{\kappa}(E_{+}/U) =  \lim_{E_0\to \infty} e^{\omega \tau}  E_{+}/( 4 E_0 ) \, . \label{eq:PhiDef}
\end{align}
The factors in Eq.~\eqref{eq:PhiDef} are chosen to set the asymptotic values of $\Phi_{\kappa}$ in the non-interacting limit, $U \to 0$,  to  $\Phi_{\kappa}(\pm \infty)=\pm 1$.
We can also interpret
$\Phi_{\kappa}$ as the exponential of the interaction-induced change in the travel time, $
   | \Phi_{\kappa}| = \exp [{\omega(\tau-\tau_{U\!=\!0})} ]$ as $\tau \to \infty$.

\begin{figure*}
    \centering
    \includegraphics[width=0.75\textwidth]{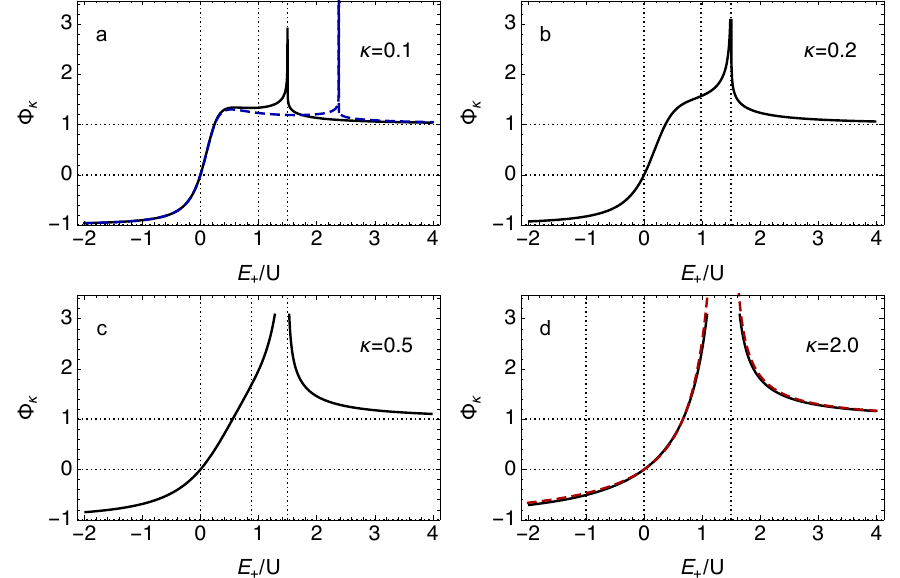}
    \caption{Scaling function $\Phi_k$ that encodes interaction-induced change in the relative coordinate travel time $\tau$. From (a) to (d): $\kappa=0.1, 0.2, 0.5$, and  $2.0$, respectively. Vertical gridlines indicate $E_{+}=0, E_{c2}$, and $E_c$.  Blue dashed line in (a): calculation with $\kappa'=\Xi \, \kappa =0.05$ and $U'=\Xi^{-2/3} U=1.587 \, U$; Red dashed line (d): $\kappa=5$.}
    \label{fig:phiplots}
\end{figure*}

The function $\Phi_{\kappa}$ for a range of $\kappa$ is shown in Fig.~\ref{fig:phiplots}.
The singularity due to critical trajectories is fixed at $E_{+}/U=3/2$ value, but the overall shape of $\Phi_{\kappa}(E_{+}/U)$ depends on the geometric parameter $\kappa= \omega_x/\omega_y$.
For large $\kappa \gg 1$ (wide constriction), the function $\Phi_{\kappa}$ converges to a $\omega_x$-independent limit as the travel time becomes interaction-dominated in a wide range of $E_{+}>E_{c2} \to -U \kappa^2/2$, see Fig.~\ref{fig:phiplots}d.  The opposite limit of small $\kappa \ll 1$ (narrow constriction) corresponds to the 1D crossover due to the suppression of motion along $y$. For energies $E_{+} <E_{c2} \approx U < E_{c} =3 U/2$, the corresponding travel time as a function of $E_{+}$ scales with  $U_{\text{1D}}=U \kappa^{2/3}$ which is a well-defined measure of the interaction strength in the 1D limit, see Eq.~\eqref{eq:U1D}. Fig.~\ref{fig:phiplots}a illustrates this scaling.

\subsection{Solution for the absolute coordinates}
We now combine the solution for the relative motion with that of the centre-of-mass motion.
From $x(0)=-x_0$ and the initial conditions \eqref{eq:timedelay}--\eqref{eq:EplusIni}, it follows that  
$x_{\text{c.m.}}(0)=(1/2) \, x_0 \tanh (\omega \, \Delta t/2)$ and (again to the first order in $E_{-}/E_0$)
\begin{align} 
    y_{\text{c.m}.}(0)&  =\kappa \, x_{\text{c.m}}(0)- \frac{\kappa \, x_0 }{ 4 E_0} E_{-} \, . \label{eq:ycminicond} 
\end{align}

In the large-$\tau$ limit the solution to the center-of-mass e.o.m.'s for the initial condition \eqref{eq:ycminicond} is
\begin{multline}
    x_{\text{c.m.}}(\tau)= - y_{\text{c.m.}}(\tau)/\kappa =e^{\omega \tau} \left [ x_{\text{c.m.}}(0)- y_{\text{c.m.}}(0)/\kappa \right]/2  \\
    = x_0 \, e^{\omega \tau} \, E_{-}/(8 E_0) \, .
\end{multline}
Combining this with the definition of $\tau$, Eq.~\eqref{eq:taudef} and $\Phi_{\kappa}$, Eq.~\eqref{eq:PhiDef},
gives the asymptotic position of the two particles after scattering,
\begin{align} \label{eq:mainresult}
\lim_{E_0 \to \infty}
 2 \, x_{j}(\tau)/x_0 = \pm 1 + \frac{E_{-} \, \Phi_{\kappa} }{E_+} \, ,
\end{align}
where the upper sign is for $j=1$ and $E_{+}<E_c$, and for $j=2$ and $E_{+}>E_c$. 

When both electrons go to the same detector ($x_1 \cdot x_2 >0$), we can express the result \eqref{eq:mainresult} as the difference in times of arrival, $\exp (\omega \Delta t^f) = \lim\limits_{t\to \infty} x_2/x_1$,
\begin{align} \label{eq:tf}
    \Delta t^{f} = \pm \, \omega^{-1} \ln \frac{ E_{-} \, \Phi_{\kappa} -E_{+} } { E_{-} \, \Phi_{\kappa} +E_{+} }  \, ,
\end{align}
(upper sign for $E_{+} < E_c$). 

The non-interacting limit of Eq.~\eqref{eq:tf}, 
\begin{align} \label{eq:tf0}
    \Delta t^{f}_{U=0}=\Delta t + \omega^{-1} \ln (-\epsilon_2/\epsilon_1) \, ,
\end{align}
reveals an energy-dependent time-shift introduced independently on each electron by the beamsplitter (classical dispersion).
Note, that the requirement $x_1 \cdot x_2 >0$ leads to $-\epsilon_2/\epsilon_1 >0$, as either electron 1 is transmitted ($\epsilon_1>0$) and electron 2 is reflected ($\epsilon_2<0$) or the other way round.
Logarithmic dependence on energy in \eqref{eq:tf0} is a direct consequence of the parabolic approximation.

\section{Phase diagram for scattering outcomes\label{sec:mainresults}}
\subsection{Phase diagram in invariant coordinates}
Classically, the scattering outcomes are deterministic unless the final state is of unstable equilibrium with one electron stuck at the saddle point. Hence the boundaries of the regions with well-defined scattering outcomes will be given by the $x_0 \to\infty$ limit of 
Eq.~\eqref{eq:mainresult} with either $x_1$ or $x_2$ finite (and hence necessarily zero). The corresponding conditions are conveniently expressed in terms of the function $\Phi_{\kappa}$ and the variables $E_{\pm}$,
\begin{align} \label{eq:boundaries}
    E_{-} \,  \Phi_{\kappa}(E_{+}/U) = \pm E_{+} \, .
\end{align}

\begin{figure*} 
    \centering
   \includegraphics[width=0.8\textwidth]{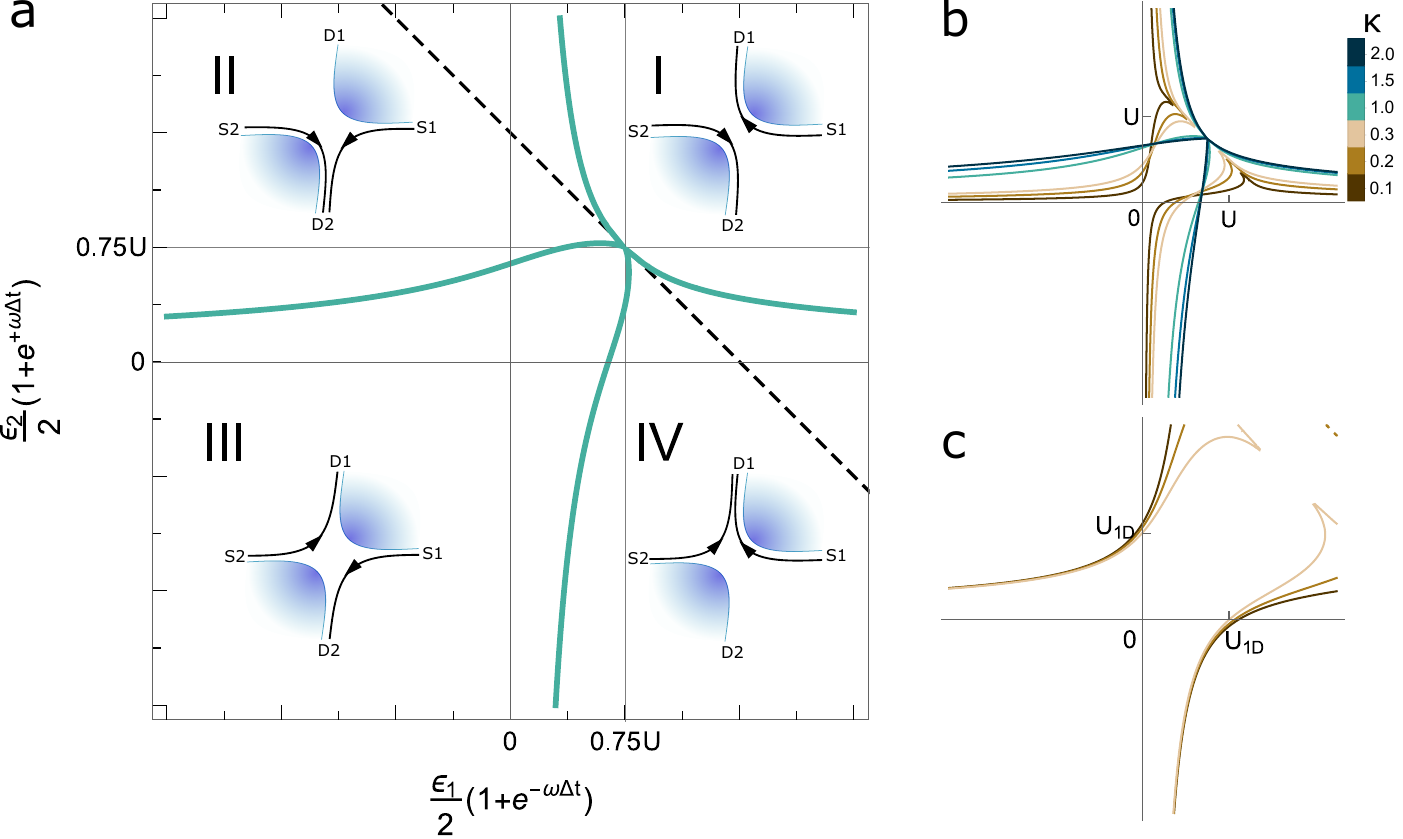} \,  
    \caption{Phase diagram of classical collision outcomes. (a) Boundaries between the four regions corresponding to well-defined number of electrons reaching the detectors: two at D2 (II), two at D1 (IV), or one at both D1 and D2 (I and III). The difference between I and III is the topology of the closest approach, as indicated by the sketches. Here $\kappa=1.0$. The dashed line is $E_{+}=E_c$. (b) Evolution of the phase diagram with changing the aspect ratio parameter $\kappa$. Coordinate axes are the same as in (a).
    In the wide constriction limit, $\omega_y \ll \omega_x$,  the diagram remains close to the case of $\kappa=2.0$ as shown  by the darkest line.
    (c). The same boundary lines as in (b), emphasizing the  narrow constriction ($\kappa=0.1$--$0.3$) limit by scaling the coordinates with $U_{\text{1D}}=U \kappa^{2/3}$ instead of $U$.} 
    \label{fig:phasediagram}
\end{figure*}

Equations \eqref{eq:boundaries} and \eqref{eq:tf}, and their subsequent analysis constitute the main result of this paper.
Separation of $E_{+}$ and $E_{-}$ variables, each given in Eq.~\eqref{eq:EplusIni} by the sum and the difference of a particular combination of the initial conditions, $\epsilon_j (1+e^{\pm \omega \Delta t})/2$, suggests  a convenient form for the phase diagram of scattering outcomes as presented in Fig.~\ref{fig:phasediagram}.  The diagram is symmetric with respect to exchange of sources S1 and S2,  $\epsilon_1 \leftrightarrow \epsilon_2$ and $ \Delta t \to - \Delta t$, due to inversion symmetry of the constriction assumed by the quadratic saddle approximation. The diagram is easiest to interpret for $\Delta t=0$ when the axes are simply the energies of the incoming electrons, $(\epsilon_1, \epsilon_2)$.

The phase diagram separates the parameter space into four domains by the topology of the connection between the incoming and the outgoing asymptotes of distinguishable electron trajectories, numbered from I to IV, as indicated by the sketches in Fig.~\ref{fig:phasediagram}a. 
The four region boundaries meet at $\{ 3U/4, 3U/4  \}$.
In regions II and IV both electrons end up in the same detector (D2 or D1 in Fig.~\ref{fig:setup}a, respectively), hence the relative time of arrival formula \eqref{eq:tf} is applicable (neglecting additional dispersion between the barrier region and the detector). The final time difference $\Delta t^f$ diverges at the boundaries \eqref{eq:boundaries} as either electron 1 or 2 remains stuck in unstable equilibrium. The dashed line marks $E_{+}=E_c$ where the function $\Phi_{\kappa}$  diverges and $\Delta t^f=0$.
Hence in region II below the dashed line and region IV above the dashed line electron $1$ arrives at the detector faster than electron $2$, $\Delta t^f<0$, and the order of arrival reverses whenever the dashed line is crossed. 

Limiting cases of the phase diagram are straightforward to interpret. The meeting point of the four regions in terms of incoming electron energies corresponds to $\epsilon_{1,2}^{\ast} =(3U/4)/[1+\exp(\mp \omega \Delta t)]$. 
If $U$ is reduced to zero, this point shifts to the origin and the diagram becomes trivial: a crossing of two uncorrelated transmission thresholds, $\epsilon^{\ast}_{1,2}=0$. For finite $U$ but large $\Delta t$, electron 2 with energy $\epsilon_2\approx\epsilon_2^{\ast} \to 0$ arrives first and ``waits'' at the constriction for electron 1. Only if the energy $\epsilon_1$ of the latter is larger than $\epsilon_{1}^{\ast} \to 3 U/2>0$ will it be sufficient not only to kick electron 2 back towards detector D2 but also for electron 1 to become transmitted to D2 (region IV) instead of being reflected to D1 (region III).

The shape of the phase diagram according to Eq.~\eqref{eq:boundaries} is completely determined by the function $\Phi_{\kappa}(E_{+}/U)$ which depends on the constriction geometry parameter $\kappa =\omega_x/\omega_y$, as already discussed in Section \ref{eq:relsol}. In Fig.~\ref{fig:phasediagram}b we show the phase  boundaries for different values of $\kappa$, using the same coordinates as in Fig.~\ref{fig:phasediagram}a.
In a wide-constriction limit, $\kappa \gtrsim 1.5$, the shape of the diagram becomes $\kappa$-independent as there is only one energy scale, $U$, that controls the collision. For $\kappa \lesssim 0.5$ an inflection point in $\Phi_{\kappa}$ develops near $E_{+} \approx E_{c2}$, and the phase diagram for $\kappa \ll 1$ shows two characteristic behaviors: (i) narrowing of the singularity at the four-region meeting  point (which is pinned on the scale of $U$), and (ii) regions II and IV approaching each other on the scale of $U_{\text{1D}} \ll U$ near the origin. The latter effect is illustrated in Fig.~\ref{fig:phasediagram}c.

In terms of non-rescaled coordinates, 1D behavior  requires not only $| \epsilon_1|, |\epsilon_2| \sim U_{1D} \ll U$ but also for the collision to take place sufficiently close to the centre of the narrow constriction.
We can get the corresponding condition on $\Delta t$ by requesting $|\epsilon_1^{\ast}|,| \epsilon_2^{\ast} | \gg U_{\text{1D}}$ which gives
$| \Delta t| \ll  -(2/3) \ln \kappa$. In the exact 1D limit, described by  $\kappa \to 0$ and $U\to \infty$ with finite $U_{1D}$ and $\omega=\omega_x$,
the region I of the phase diagram does not exist and the limiting form illustrated approximately Fig.~\ref{fig:phasediagram}c with $\kappa=0.1$  becomes universal for one-dimensional Coulomb scattering.

\subsection{Collision outcomes in experimentally relevant variables\label{sec:experimental}}
The universal phase diagram in Fig.~\ref{fig:phasediagram} can be explored by scanning  different combinations of experimentally controllable parameters,  $\epsilon_1$, $\epsilon_2$ and $\Delta t$. Here we examine a particular protocol~\cite{MasayaPrivate}: changing the average energy  $\epsilon_0 =(\epsilon_1+\epsilon_2)/2$ and the relative delay time $\Delta t$, while keeping the energy mismatch between the sources S1 and S2 constant, $\Delta \epsilon = \epsilon_1-\epsilon_2=\text{const}$. Note that 
changing $\epsilon_0$ is equivalent to gating the whole saddle point region~\cite{Locane2019} (varying the scattering barrier height).

  \begin{figure}
    \centering
   \includegraphics[width=0.45\textwidth]{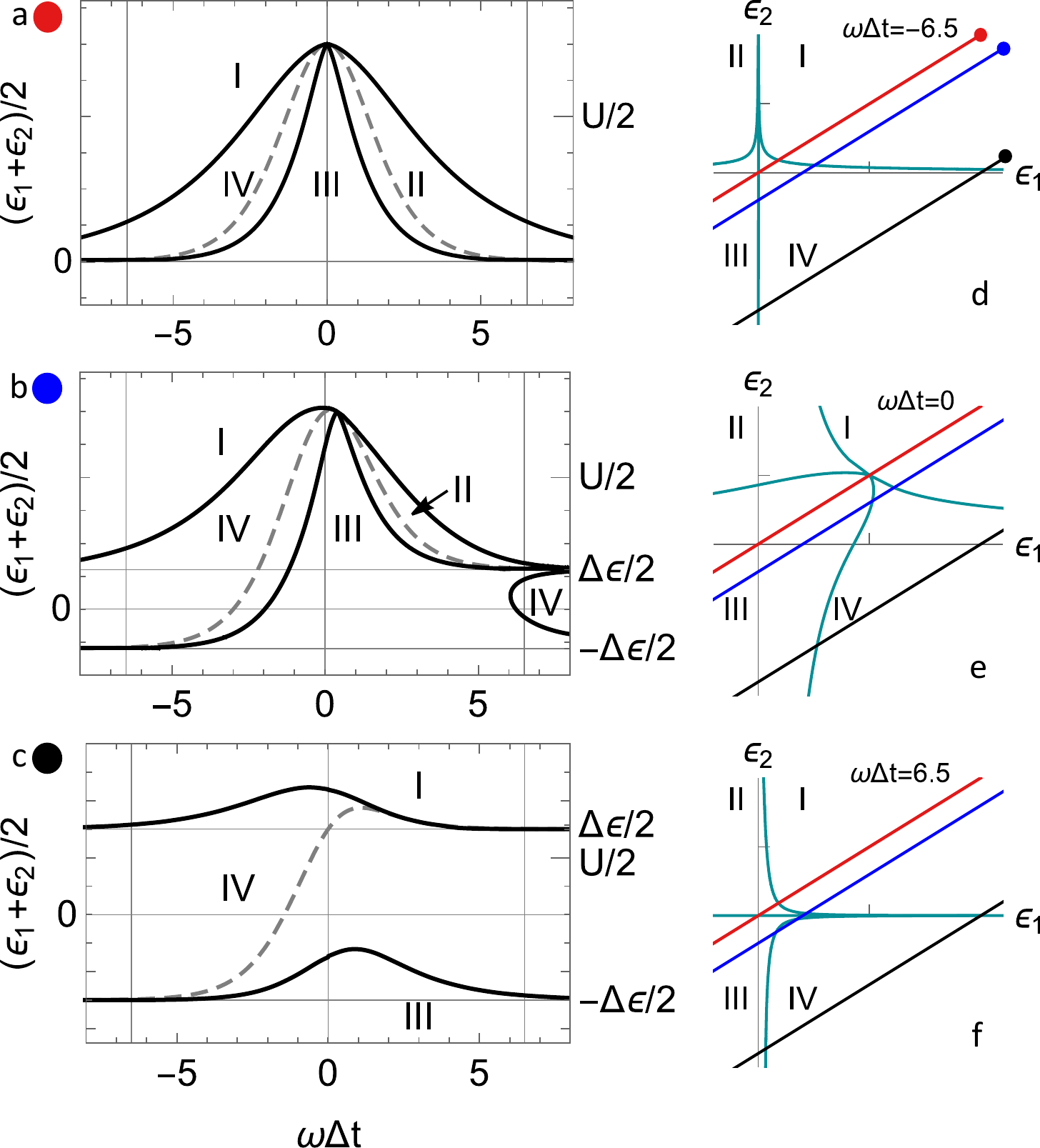}
    \caption{Predicting scattering outcomes as a function of average electron energy and relative time for (a) zero, $\Delta \epsilon =0$, (b) intermediate, $\Delta \epsilon = 0.3U$, and (c) large, $\Delta \epsilon = 1.5U$, energy differences between the electrons. The dashed line indicates $E_{+}=E_c$. Panels (d) to (f) show cuts though the phase diagram drawn in terms of $\epsilon_1$ and $\epsilon_2$ at particular fixed values of $\Delta t$ (quoted in the diagrams) and at the values of $\Delta \epsilon$ fixed in panels (a) to (c). Ticks on the axes in (d) to (f) mark the value of $0.75\, U$. Colored diagonal lines in (d) to (f) trace vertical cuts shown in graphs (a) to (c), with correspondence indicated by red, blue and black dots, respectively. Aspect ratio parameter is $\kappa = 1$.}
    \label{fig:NPLstyle}
\end{figure}

In Figure \ref{fig:NPLstyle} we show three examples corresponding to  zero, intermediate, and large $\Delta \epsilon$ in panels (a) to (c), respectively.
The sketches in Fig.~\ref{fig:NPLstyle}d-f show the phase diagram in $\{ \epsilon_1, \epsilon_2\}$ coordinates for three fixed values of $\Delta \epsilon$. Following the cuts of constant $\Delta \epsilon$ and $\Delta t$ in $\{ \epsilon_1, \epsilon_2\}$ plane [marked by colored lines in panels (d)--(f)] reveals the sequence of scattering outcomes along the corresponding vertical cuts in diagrams (a)--(c).

The diagram for equal incoming energies is shown in Fig.~\ref{fig:NPLstyle}a.
If the particles arrive simultaneously,  $\Delta t \approx 0$, they go to opposite detectors, either both passing through (region I) or getting reflected from the barrier and one another (region III). Such perfect anti-correlation would be detectable as a suppression of crosscorrelation noise between D1 and D2.   The regions II and IV (``wings'' of the diagram) are characterized by an unequal distribution of the current between the detectors, and can be distinguished by a differential directed current measurement between D1 and D2. In region II,  both particles end up in D1 if electron 2 arrives first  ($\Delta t >0$), i.e.\ electron 2 is transmitted and electron $1$ is reflected (see the sketch for region II in Fig.~\ref{fig:phasediagram}), even though the barrier height should allow transmission of electron 2 in the absence of the other electron ($\epsilon_1 =\epsilon_2 >0$ in regions II and IV). 

A similar shift in a transmission threshold  towards higher incoming energies is observed if the electron energies are not equal, see Fig.~\ref{fig:NPLstyle}c.  Near coincidence ($\omega \Delta t \sim 1$) one observes ``bumps'' in otherwise horizontal boundary lines at $\epsilon_0 = \pm \Delta \epsilon/2 \Leftrightarrow \epsilon_{1,2}=0$.
This increase in effective  barrier height due the Coulomb repulsion by the other electron is quantified in our model by $U$ and $\omega$, and can be tested even if the fourfold degeneracy point $\epsilon_j=\epsilon_j^{\ast}(\Delta t)$ is not reached or is confounded by broadening effects. 

For $0< |\Delta \epsilon| < 3 U /2$, complex intermediate cases are possible, as exemplified in Fig.~\ref{fig:NPLstyle}b for $\Delta \epsilon =0.3 U$.

\section{Conditions for the classical limit \label{sec:applicability}}


We now discuss consistency conditions for the classical solution that will help us to estimate the boundaries in the parameter space where the solution is a valid approximation.

\subsection{Quantum broadening\label{sec:hbaromega}}
So far we have treated the electrons as point particles that can have a well-defined energy at a well-defined time. Inevitable uncertainty due to quantum mechanics (and potential additional classical fluctuations at the source~\cite{Fletcher2019}) will result in probabilistic scattering and broaden the sharp lines of the phase diagram discussed in the previous Section. 

A qualitative condition in the energy domain for applicability of the classical picture follows from the exact solution of the single-particle quantum scattering on the saddle point potential~\cite{Fertig1987} (summarized in Appendix~\ref{sec:quantumsp}).
Transmission is near-deterministic (probability close to $0$ or $1$) and the travel times computed from the group velocity of a wave-packet follow closely the classical equation Eq.~\eqref{eq:nonint} if the energy distance $E$ to the saddle point is larger than $\hbar \omega$.
This condition is immediately applicable to the centre-of-mass degree of freedom since it is governed by the same potential as a single particle,
\begin{align} \label{eq:quantumCM}
|E_{\text{c.m.}}| = | E_{-} \tanh ( \omega \Delta t/2) | > \hbar \omega \, .
\end{align}
The condition \eqref{eq:quantumCM} is independent of $U$ and reflects the quantum uncertainty of coincidence in both time and energy. At $\omega \Delta t \lesssim 1$, Eq.~\eqref{eq:quantumCM} implies
\begin{align}
    (1/2) | (\epsilon_2-\epsilon_1) \, \Delta t | > \hbar \, ,
\end{align}
as expected from the uncertainty principle applied to each wave-packet individually before scattering.

For the relative motion, the relevant saddle point is given by the quadratic expansion of Eq.~\eqref{eq:cons2} near $(0, -d_0)$,
\begin{align} 
   E_{+}(x,y)-E_c \approx  m  \left [
   3\, \omega_y^2 \, (y+d_0)^2 - (\omega_x^2+\omega_y^2) \, x^2 \label{eq:saddleexp}
   \right ]/4 \, . 
  \end{align}
  Note that the expansion Eq.~\eqref{eq:saddleexp} is valid only for trajectories that approach the saddle-point ---  this requires at least $E_{+} > E_{c2}$,
see Fig.~\ref{fig:levelsrel}. Comparing Eq.~\eqref{eq:saddleexp} to Eq.~\eqref{eq:cons2} with $d_0=0$ we see that instead of $\omega =\omega_x \omega_y/\omega_c$, the vicinity of the  interaction-induced saddle point is controlled by the time scale
\begin{align}
    \omega^{(2)} = \omega_y \sqrt{3 (\omega_x^2  +\omega_y^2)}/\omega_c
\end{align}

and the corresponding condition is
\begin{align}
    |E_{+} - (3 \, U/2) | > \hbar \omega^{(2)} =\hbar \omega \sqrt{3(1+\kappa^{-2})} \, .\label{eq:eq:quantumrel}
\end{align}
This condition will necessarily be violated near the dashed lines in the diagrams of Figures~\ref{fig:phasediagram} and \ref{fig:NPLstyle}, and in particular, near the fourfold degenerate point. Yet for a sufficiently large $U$ the bulk areas of the diagram will be robust against quantum uncertainty if the colliding wave-packets are prepared sufficiently compact in energy and time.

\subsection{Consistency conditions for the classical approximation}
If the interaction strength is insufficient, the quantum effects will always be overwhelming and the classical description of interactions will fail qualitatively.  Hence it is important to estimate the lower bound on $U$ due to quantum mechanics. (There is also an upper bound imposed by the condition of staying in the lowest confined mode, discussed further below).

For moderate and wide constrictions, $\kappa \gtrsim 1$, $U$ is the only energy scale in the classical phase diagram, and $U$ has to be larger than the quantum broadening scale \eqref{eq:eq:quantumrel},
\begin{align} \label{eq:Ucond2D}
    U \gg \hbar \omega \, .  \quad \text{(wide)} 
\end{align}
This condition has a straightforward physical interpretation in terms of the experimental protocol discussed in Section~\ref{sec:experimental}: the increase in the effective barrier height due to the presence of another electron, $U$, has to be larger than the barrier energy resolution, $\hbar \omega$, to allow for a single-shot detection of coincident arrival.

For a narrow constriction, $\kappa \ll 1$, there are two behaviours: 2D-like in the vicinity of the fourfold degenerate point, for $E_{+} \sim U$, for which the condition \eqref{eq:eq:quantumrel} implies
\begin{align} \label{eq:Ucond2Dnarrow}
    U \gg \hbar  \omega_y^2/\omega_c \, , \quad \text{(narrow)}
\end{align}
and the 1D-like behaviour for $|E_{+}| \sim U_{1D} \ll U$. 
Together with the microscopic definitions of $U=(m/2)^{1/3} \, [\omega_y \, e^2/(4 \pi \varepsilon_0 \varepsilon) ]^{2/3}$ and $\omega=\omega_x \, \omega_y/\omega_c$,
conditions \eqref{eq:Ucond2D}--\eqref{eq:Ucond2Dnarrow} define the necessary bounds for confinement strengths $\omega_x, \omega_y \ll \omega_c$ in a material with known effective mass $m$ and dielectric constant $\varepsilon$.

The conditions \eqref{eq:Ucond2D} and \eqref{eq:Ucond2Dnarrow} expressed in energy language can also be understood in terms of phase-space geometry of Fig.~\ref{fig:levelsrel}. 
The area $A$ of the inaccessible region at short relative distances, enclosed by the critical level line $E_{+}=E_c$ in Fig.~\ref{fig:levelsrel}, has to be much larger than the quantum phase space unit $l_c^2$. A straightforward computation gives $A \sim d_0^2/\kappa$  for $\kappa \gtrsim  1$, and  $A \sim d_0^2$ for $\kappa \to 0$. The condition $A \gg l_c^2= \hbar/(m \omega_c)$ is then equivalent to either \eqref{eq:Ucond2D} or \eqref{eq:Ucond2Dnarrow} for the respective range of $\kappa$.

The classical solution completely 
neglects the effects of quantum statistics. Qualitatively, this can be similarly justified by the localization length $l_c$ of maximally localized quantum wave packets being smaller than the minimal distance allowed by interactions (which is derived in the classical limit in Appendix~\ref{sec:ksiplot}), yet a careful analysis of potential quantum exchange effects in the strong-coupling limit is beyond the scope of this study.

A bound on $U$ from above follows from the condition of scattering within the first Landau level only,
\begin{align} \label{eq:condUpper}
    \hbar \omega , \, U \ll \sqrt{\omega_y^2+\omega_c^2}  \, .
\end{align}
This can be satisfied in large magnetic fields for $ \omega_x , \omega_y \ll \omega_c$ both for a wide or a narrow constriction.  If, on the other hand, the electrostatic transverse confinement is significant, $\omega_y \gtrsim \omega_c$, then only the narrow limit is allowed, $\omega_x \ll \omega_y$, but then Eq.~\eqref{eq:condUpper} becomes incompatible with the condition for the classical 2D behavior \eqref{eq:Ucond2Dnarrow}. Hence we explicitly confirm that magnetic confinement is essential for the possibility to probe all  four regions of the classical phase diagram.






\section{Conclusions and outlook\label{sec:conclusions}}

The classical-limit phase diagram of two-electron Coulomb scattering in two dimensions, described in Section~\ref{sec:mainresults}, is a robust and a tightly constrained prediction since it maps a three-dimensional manifold of initial conditions onto a two dimensional diagram of final outcomes which is fully determined \emph{ab initio}. Scaling with particular combinations of initial conditions ($E_{\pm}$) is potentially testable experimentally even in the presence of significant stochastic broadening.

In addition to measuring collision outcomes, time-of-flight measurements \cite{Kataoka2016a} could be  used to characterize the classical dispersion of the constriction. An additional element is a gate-controlled  ``chopper'' barrier in front of a detector that is triggered at a tuneable time delay with respect to the source(s)~\cite{Fletcher2019}.  
On the single-particle level,
one could calibrate Eq.~\eqref{eq:nonint} [Eq.~\eqref{eq:quantumtime}] for a single source or  
Eq.~\eqref{eq:tf0} for two sources. Such classical partitioning (i.e., deflection) measurements would yield not only an estimate of $\omega$ but also of the range of $\Delta t$ and $\epsilon_j$ for which a quadratic saddle point approximation is applicable. 
Experimental techniques have already been demonstrated to resolve the time gap in the arrival of two electrons at one detector~\cite{Waldie2015}, thus our quantitative prediction for $\Delta t^{f}$, Eq.~\eqref{eq:tf}, could also be put to the test alongside with the  diagram of scattering outcomes.

The classical approach to electron scattering presented here follows the spirit of classical interpretation \cite{Kataoka2016pss} of energy-time tomography of isolated on-demand electrons demonstrated recently by Fletcher \emph{et al.}~\cite{Fletcher2019}. In both cases, fidelity of the outcome improves with reducing the characteristic scale $\hbar \omega$  for energy sensitivity of tunneling\footnote{Usually denoted $2 \pi \Delta_b$ for tuneable-barrier devices~\cite{Ubbelohde2015,Kaestner2015,Waldie2015}.}, as compared to the interaction strength $U$ in our case and the energy width of the incoming distribution $\sigma_E$ in the case of tomography. This is \emph{opposite} to the HOM-interference-based tomography of low-energy excitations close to a Fermi surface~\cite{Jullien2014,Bisognin2019} that works with spectrally neutral half-transmission beamsplitters on the energy scale much \emph{smaller} than $\hbar \omega$~\cite{Bocquillon2013}. Exploration of the crossover between these two extremes of electron-electron collisions presents a challenging non-perturbative problem for theory. 

Yet another closely related experimental system in which developing a classical approach to scattering similar to the present study could be potentially useful is the on-demand transport of electrons in potential minima induced by a travelling surface acoustic wave (SAW)~\cite{Bertrand2016,Takada2019}. There, a single-electron beamsplitter has been recently  realized~\cite{Takada2019} and time-of-flight measurements have been demonstrated~\cite{Edlbauer2021} which potentially would allow one to bring two electrons simultaneously to the interaction and tunneling region from independent sources. Energy scales analogous to our $U$ and $\hbar \omega$ could play a comparable role for determining the physical regime of two-particle collision in such SAW devices, and estimates of a sizeable phase space available at the beamsplitter~\cite{Takada2019} suggest room for suitable classical approximations.

We hope that the results of this study offer a useful map
for a particular corner of strongly interacting few-electron mesoscopic systems ripe for exploring novel fundamental effects \cite{Silvestrov2022} and developing technology for applications~\cite{Johnson2016}.


\acknowledgments
We thank Niels Ubbelohde, Piet Brouwer, Masaya Kataoka, and Martins Kokainis for discussions.
EP, GB and VK have been supported by the Latvian Council of Science (project no.\ lzp-2020/2-0281). PS and PR acknowledge financial support by the Deutsche Forschungsgemeinschaft (DFG, German Research Foundation) within the framework of Germany’s Excellence Strategy-EXC-2123 QuantumFrontiers-390837967. EP acknowledges additional support of University of Latvia foundation sponsored by University of Latvia patron ``Mikrotīkls''.
This work was supported in part by the Joint Research Project SEQUOIA (17FUN04) which received funding from the
European Metrology Programme for Innovation and Research (EMPIR) cofinanced by the Participating States and from the European Unions Horizon 2020 research and innovation programme.

\appendix
\section{Details of exact single-electron solution\label{sec:FH} and derivation of one-dimensional equations of motion.}
Here we give details of the exact quantum solution to a single particle in a saddle-point potential~\cite{Fertig1987},
and provide a first-principles derivation of the classical equations of motion \eqref{eq:vDrfit} and the scaling relations \eqref{eq:rescaled}. 
\subsection{Exact diagonalization\label{sec:FHstart}}
Fertig and Halperin \cite{Fertig1987}
put the quadratic single-particle 
Hamiltonian \eqref{eq:Hsp} into a diagonal form\footnote{We omit the electron index $j=1,2$ since only a single-particle problem is discussed throughout this Appendix.}
\begin{align} \label{eq:FHhamiltonian}
    \mathcal{H} = \frac{\hbar \omega_2}{2} (s^2 + p^2) +
    \frac{\hbar \omega_1}{2} (P^2-X^2)
\end{align}
by a linear transformation to new separated canonically conjugate variables such that $[X,P]=[s,p]=i$ and 
$[s,X]=[s,P]=[p,X]=[p,P]=0$ (in their solution $\omega_c>0$ corresponds to $\bm{B}$ in the positive direction of $z$ axis). 

The two frequencies $\omega_2$ and $\omega_1$ are given by the positive solutions to 
\begin{subequations}
\label{eq:E1E2relations}
\begin{align}
   \omega_1 \omega_2 & = \omega_x \omega_y \\
   \omega_2^2 -  \omega_1^2  \, , &  = \omega_c^2-\omega_x^2+\omega_y^2 \, .
\end{align}
\end{subequations}

Considering $\omega_2 > \omega_1$,
denoting $\eta =\omega_1 / \omega_2$, and introducing the confinement length $l_2=\sqrt{\hbar / (m \,  \omega_2)}$, we cast the exact diagonalization transformation of Ref.~\onlinecite{Fertig1987} into the following form (as in the main text, we denote $\kappa= \omega_x/\omega_y$):
\begin{subequations}
\label{eq:exacttransform}
\begin{align}
    x/l_2 & = \sqrt{\frac{\kappa^{-1} +\eta}{1+\eta^2}} \, X  - \sqrt{\frac{1- \kappa^{-1} \, \eta}{1+\eta^2}}  \, s \label{eq:xtoXP}
    \\
    y/l_2 & = \sqrt{\frac{\kappa -  \eta}{1+\eta^2}}  \, P  + \sqrt{\frac{1+ \kappa \, \eta }{1+\eta^2}}  \, p \label{eq:ytoXP}
    \\
    l_2  \, p_x /\hbar & =\frac{\kappa+\eta}{2} 
   \sqrt{\frac{\kappa^{-1} +\eta}{1+\eta^2}} \,  P - \frac{1-\kappa \, \eta}{2} 
   \sqrt{\frac{1- \kappa^{-1} \, \eta}{1+\eta^2}}  \, p
  \label{eq:pxFH} \\
   l_2 \, p_y /\hbar & =
   -\frac{\kappa^{-1}-\eta}{2}  
  \sqrt{\frac{\kappa- \eta}{1+\eta^2}} 
  \, X  
  - \frac{1 + \kappa^{-1} \eta}{2} \sqrt{\frac{1+ \kappa \, \eta }{1+\eta^2}} \,  s
\end{align}
\end{subequations}
Instead of dimensionless $X$ and $P$ of Ref.~\onlinecite{Fertig1987}, we work with dimensionful variables $\tilde{x} \propto X$ and $\tilde{y} \propto P$, defined by setting $s \to 0$, $x \to \tilde{x}$ ,
$p \to 0$,   and $y\to \tilde{y}$ in Eqs.~\eqref{eq:xtoXP} and \eqref{eq:ytoXP},
\begin{subequations} \label{eq:transform}
  \begin{align}
    x & = \tilde{x} - s \, l_0^{x} \, , \\
    y & = \tilde{y} + p \, l_0^{y} \, ,
    \end{align}
  \end{subequations}
 with some $l_0^x$ and  $l_0^y$ with the dimension of length.
 
 The commutation relation between $\tilde{x}$ and $\tilde{y}$ follows from $[X,P]=i$ and Eqs.~\eqref{eq:transform}, 
\begin{align} \label{eq:commutationElegant}
[\tilde{x},\tilde{y}] = i \, l_0^{x} \, l_0^{y} \, .
\end{align}
Note that $[\tilde{x},s]=[\tilde{y},p]=
[\tilde{x},x]=[\tilde{y},y]=0$.

\subsection{Separation of scales\label{eq:separation}}

In the limit of $0<\eta = \omega_1 / \omega_2  \ll 1$, it follows from Eqs.~ \eqref{eq:E1E2relations} that   $\omega_2 \approx \sqrt{\omega_c^2+\omega_y^2-\omega_x^2}$ and 
\begin{align}
   \eta 
  \approx \frac{\omega_x \, \omega_y}{\omega_c^2+\omega_y^2-\omega_x^2} \ll 1 
 \label{eq:separationCondition}
\end{align}
It is easy to deduce from Eq.~\eqref{eq:separationCondition} that  separation of scales, $\omega_1 \ll \omega_2$,  implies $\omega_x \ll \max( \omega_c, \omega_y)$, that is, either strong magnetic ($\omega_c$) or electric ($\omega_y$) confinement.

To the leading order in $\eta$, we have $\{ \omega_x, \omega_y, \omega_c\} /\omega_2 \approx \{ \sqrt{ \eta \, \kappa} , \sqrt{ \eta / \kappa} , \sqrt{1- \eta/\kappa} \}$
 (note that $\kappa$ can be of order $\eta$ if $\omega_y \gg \omega_c$) and 
 \begin{subequations}
 \label{eq:omegaFH} 
\begin{align}
    \omega_1 & = \frac{\omega_x \, \omega_y}{\omega_2} \, ,\\
    \omega_2 & \approx \sqrt{\omega_y^2+\omega_c^2} \, .
\end{align} 
\end{subequations}
Equations \eqref{eq:omegaFH} justify the formulas for $\omega$ and $\omega'$ used in the main text in and before Eq.~\eqref{eq:rescaled}. 

The transformation \eqref{eq:exacttransform} simplifies to 
\begin{subequations}
\begin{align}
    x/l_2 & = \kappa^{-{1/2}} \left ( X - \sqrt{\kappa- \eta}\,s  \right ) \\
   y/l_2 & = \sqrt{\kappa- \eta} \, P + p   \\
    l_2 p_x/ \hbar & =2^{-1} \, \kappa^{-{1/2}} \left [ (\kappa +\eta) \, P - \sqrt{\kappa- \eta}  \, p \right ]  \label{eq:pxsimple} \\
   l_2 p_y/ \hbar & =2^{-1} \, \kappa^{-{1}} \left [ -\sqrt{\kappa- \eta}  \, X - (\kappa +\eta) \, s \right ] \end{align}
  \end{subequations}
and the corresponding characteristic lengths in Eqs.~\eqref{eq:transform} 
become simply $l_0^{x}=l_2 \, \omega_c/\omega_2$ and  $l_0^{y}=
 l_2$.

The same limit of $\eta \to 0$ also simplifies the Hamiltonian of the propagating dimension in Eq.~\eqref{eq:FHhamiltonian},
    \begin{align} \label{eq:Ham1Dtilde}
    \frac{\hbar \omega_1}{2} (P^2-X^2) & = V_{\text{saddle}}(\tilde{x},\tilde{y}) +\frac{\omega_y^2}{\omega_c^2} \times \frac{m \omega_y^2 \, \tilde{y}^2 }{2} \, .
\end{align}

\subsection{Reduction to one-dimensional motion of the guiding centre\label{sec:derive1D}}
Consider a particle which in addition to the saddle point potential and the magnetic field
captured by $\mathcal{H}$ from Eq.~\eqref{eq:Hsp} [\eqref{eq:FHhamiltonian}]
is subject to external potential $V(x,y)$.
For the two-body interaction problem considered in the main text $V$ is the interaction potential that also depends on the coordinates of the other particle; here we focus on the formal procedure for a generic $V(x,y)$.
It is clear from Eq.~\eqref{eq:transform} that quantum fluctuations of the confined degree of freedom $(s,p)$ introduce uncertainty to $x$ and $y$ on the scale of $l_2$. If $V(x,y)$ is smooth on this scale, we can develop a useful approximation for one-dimensional motion, assuming that $s$ and $p$ are confined to the lowest energy state (lowest Landau level/transverse quantization mode) and using $\tilde{x}$ and $\tilde{y}$  as the active coordinates for the guiding centre motion.

 Using the saddle-point Hamiltonian~\eqref{eq:FHhamiltonian},  the commutation relations  \eqref{eq:commutationElegant} and the simplifications of the $\eta \ll 1$ limit, Eq.~\eqref{eq:Ham1Dtilde} and
  $l_0^x \, l_0^y =  \hbar \omega_c/ [m(\omega_c^2+\omega_y^2)]$,  Heisenberg equations of motion for the guiding centre coordinates $(\tilde{x},\tilde{y})$ are 
 \begin{subequations}
 \label{eq:EOMs}
  \begin{align}
     \dot{\tilde{x}} & = \frac{i}{\hbar} [ \mathcal{H}+V(x,y),\tilde{x}] =  \tilde{y} \, \omega_y^2/ \omega_c + \frac{1}{m}  \frac{\partial V}{\partial y} \times \frac{\omega_c}{\omega_y^2+\omega_c^2} \\
  \dot{\tilde{y}} & = \frac{i}{\hbar} [\mathcal{H}+V(x,y),\tilde{y}] = \left[  \tilde{x} \, \omega_x^2 - \frac{1}{m}  \frac{\partial V}{\partial x} \right ] \times \frac{\omega_c}{\omega_y^2+\omega_c^2} 
  \end{align}
\end{subequations}
Coupling between ($s,p$) and $(\tilde{x},\tilde{y})$ is present in Eqs.~\eqref{eq:EOMs} due to difference between $(x,y)$ and $(\tilde{x},\tilde{y})$, but the equations are still formally exact (apart from using the separation of scales simplifications).

Tracing out $s$ and $p$ requires an assumption about the state of the confined dimension. Assuming the lowest Landau level, which corresponds to the ground state of the corresponding harmonic oscillator in Eq.~\eqref{eq:FHhamiltonian}, 
the projection can be written explicitly in the coordinate representation of $s$ and $p=- i \partial /\partial s$,
\begin{align} \label{eq:Vprojection}
    \tilde{V}(\tilde{x},\tilde{y} ) =\frac{1}{\sqrt{\pi}} 
     \int\limits^{+\infty}_{-\infty} e^{-s^2/2} V(\tilde{x} - s l_0^{x}, \tilde{y} - i l_0^{x} \partial_s) e^{-s^2/2} \, d s
\end{align}
Performing a similar projection on Eqs.~\eqref{eq:EOMs} would give Heisenberg equations of motion for position-momentum operator pair $\tilde{x}$, $\tilde{y}$ with $l_0^{x} \, l_0^{y}$ playing the role of an effective Planck constant.

The classical limit formally corresponds to $\omega_2 \to \infty$, which leads to $l_0^{x} , l_0^{y} \to 0$, $\tilde{V}(\tilde{x}, \tilde{y}) \to V(x,y)$  and turns Eq.~\eqref{eq:EOMs} into the conjugate pair of  Hamilton equations for the classical trajectory $x(t)$, $y(t)$.

Taking the limit $\omega_c \to \infty$ in  Eqs.~\eqref{eq:EOMs} and \eqref{eq:Vprojection} and identifying $m B =\hbar\omega_c$ gives\footnote{One also has to flip the overall sign of Eqs.~\eqref{eq:EOMs} 
to account for the direction of $\bm{B}$ which is parallel to $z$-axis in Ref.~\onlinecite{Fertig1987} on which Appendix~\ref{sec:FH} is based and antiparallel elsewhere in the paper.}
Eqs.~\eqref{eq:vDrfit} of the main text which are 
 simply statements  of drift velocity $\bm{v}= \nabla V \times \bm{B}/(e B^2)$ for each electron in the combined electrostatic field of external confinement and mutual repulsion.
 
\subsection{Mapping onto $E \times B$ drift for arbitrary electric-to-magnetic confinement ratio $\omega_y / \omega_c$\label{sec:map}}

 The first-principles derivation laid out in Sections~\ref{sec:FHstart}--\ref{sec:derive1D}  relies only on $\omega_x \ll \sqrt{\omega_y^2+\omega_c^2}$ for separation of scales and hence does  require  $\omega_y \ll \omega_c$ as a  necessary condition. 
We observe that the $B \to \infty$ drift velocity equations \eqref{eq:vDrfit} used to derive the results of  this paper coincide with the classical limit of Eqs.~\eqref{eq:EOMs} if $\omega_c$ and $\omega_y$ in the former are replaced by 
\begin{subequations}
\label{eq:motherofscaling}
\begin{align} 
    \omega_c' &=\omega_c + \frac{\omega_y^2}{\omega_c} = \omega_c \, \Xi^{-2}  \label{eq:omegacprime} \\
    \omega_y' &= \frac{\omega_y}{\omega_c} \sqrt{\omega_c^2+\omega_y^2} = \omega_y \, \Xi^{-1} 
\end{align}
\end{subequations}
This observation yields the rescaling recipe \eqref{eq:rescaled} of the main text.  We also note that $l_0^x= \Xi^{3/2} \, l_c$, $l_0^y= \Xi^{1/2} \, l_c$ and $[\hat{x},\hat{y}]   =i (l_c')^2$ where $l_c'=\sqrt{\hbar/( m \omega_c')}=\Xi \, l_c$ is the renormalized magnetic length.

In the 1D limit, $\omega_c/\omega_y \to 0$, we can use \eqref{eq:motherofscaling} on Eq.~\eqref{eq:pxsimple} to confirm the correspondence of operators $p_x = m \omega_y^2\, \tilde{y}/\omega_c$, consistent with $\Xi \to 0$  derivation of the Newton's second law for the relative coordinate on the classical level, as discussed after Eqs.~\eqref{eq:rescaled} in the main text.

\subsection{Single-particle quantum scattering on the saddle potential\label{sec:quantumsp}}
Quantum scattering probability on the saddle-point (in 2D terms) or parabolic (in equivalent 1D representation) potential for a wave-packet with a well-defined energy $E$ is~\cite{Kemble1935,Fertig1987,Buttiker1990}:
\begin{align} \label{eq:TofE}
  T(E) = \frac{1}{1+\exp [-2 \pi E/(\hbar \omega)]} \, ,
\end{align}
where $\omega=\omega_1$ of the exact diagonalization \cite{Fertig1987} described in Section~\ref{eq:FHhamiltonian}.
We see that $T(E)$ is exponentially close to either $0$ or $1$ (i.e.\ classical) if $|E| \gg \hbar \omega$. 

In the time domain, quantum fluctuations heal the logarithmic divergence near the saddle point on the same energy scale~\cite{SilvestrovPRE02}. The reflection (transmission) time from $x=-x_0$ to $x=-x_0$ ($x=+x_0$) computed as a Wigner delay time $\tau_{\text{W}} =\hbar^{-1} \partial \im \log s_\alpha/ \partial E$  from the asymptotically exact quantum scattering amplitudes \cite{Fertig1987,Lackenby2014} $\alpha=\mathrm{R}$ ($\alpha =\mathrm{T}$), 
\begin{subequations}
\label{eq:scatteringamplitudes}
\begin{align}
    s_{\mathrm{T}} = &  
     (2 \pi)^{-1/2} \, e^{ (0.5  \pi E+ i 2 E_0)/(\hbar \omega)} \\ \times \nonumber
     & \left (\frac{4 E_0}{\hbar \omega}\right)^{i E/(\hbar \omega)} 
    \Gamma\left (\frac{1}{2} + i \frac{E}{\hbar \omega} \right ) \, , \\
    s_{\mathrm{R}}  = &  -i e^{- \pi E /(\hbar \omega)} \, s_{\text{T}} \, , 
    \label{eq:sR}
\end{align}\end{subequations}
equals to
\begin{align}
    \omega \, \tau_{W}(E) = \ln \frac{4 E_0}{\hbar \omega} - \re \Psi\left (\frac{1}{2} + i \frac{E}{\hbar \omega} \right ) \, , \label{eq:quantumtime}
\end{align}
where $\Gamma(z)$ is the gamma and $\Psi(z)=d \ln \Gamma(z)/dz$ is the digamma function.
Equations~\eqref{eq:TofE} and \eqref{eq:scatteringamplitudes} are related by $T(E)=|s_{\mathrm{T}}|^2=1-|s_{\mathrm{R}}|^2$.

\begin{figure}
    \centering
   \includegraphics[width=0.4\textwidth]{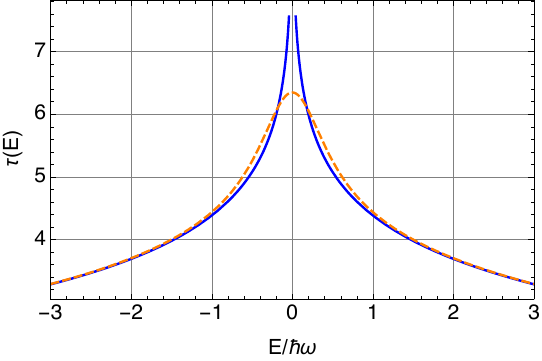}
    \caption{Continuous (blue) line: classical travel time \eqref{eq:nonint} for $E_0=20 \, \hbar \omega$, dashed (orange) line: quantum wave-packet travel time (computed as Wigner delay time with an appropriate phase reference point), both as functions of single-particle energy $E$.}
    \label{fig:delaywigner}
\end{figure}

The Wigner delay time Eq.~\eqref{eq:quantumtime} should be compared to the classical travel time \eqref{eq:nonint} with a matching phase reference  point $\pm x_0$, single-particle energy $E_{+}=E$ and mass (here $m$ and hence $E_0=m \omega_x^{2} x_0^2/2$, in contrast to the reduced mass $\mu=m/2$ and $E_0=\mu \omega_x^{2} x_0^2/2$ in Section~\ref{eq:relsol}). In the limit of low tunnelling probabilities, $|E| \gg \hbar \omega$,  as $\Psi(z) \sim \ln z $ at $|z| \gg 1$,  the quantum mechanical calculation gives the same result as the classical one; the classical divergence  is cut off at $|E| \approx \hbar \omega$ giving a finite 
$\tau_{W}(E\!=\!0)=\omega^{-1} \left[ -\ln (\hbar \omega/E_0) +3.35\ldots\right]$.
This comparison is illustrated in Fig.~\ref{fig:delaywigner}. 

\section{Minimal distance\label{sec:ksiplot}}
Here we 
evaluate the minimal classical distance between electrons as function of the variable $E_{+}$ that controls the dynamics of the relative coordinate
.
\begin{figure}
    \centering
   \includegraphics[width=0.45\textwidth]{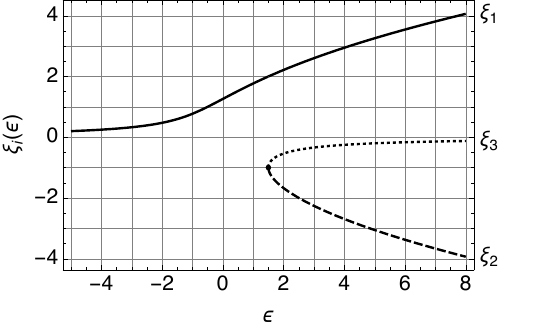}
    \caption{Roots of Eq.~\eqref{eq:cubic} for calculating the minimal distance. The degeneracy point $\xi_2=\xi_3$ is at $\{3/2,-1\}$.}
    \label{fig:ksiplot}
\end{figure}

Minimal distance $d_{\text{min}}$ between two electrons is reached at $t=\tau/2$ at the relative coordinate vector  equal to either $\{ - d_{\text{min}}, 0 \}$ for $E<E_c$ or
$\{ 0, - d_{\text{min}} \}$ for $E>E_c$, cf. Fig.~\ref{fig:levelsrel}.
The corresponding values as a function of $\kappa$ and $E_{+}$ can be expressed analytically 
 in terms of roots of the a cubic equation,
\begin{align} \label{eq:cubic}
   \xi^3 - 2 \, \epsilon \, \xi -2=0 \, .
\end{align}
The real roots of this  equation are plotted in Fig.~\ref{fig:ksiplot} as functions of $\epsilon$.
There is one positive real root $\xi_1(\epsilon) > 0$ for all real  $\epsilon$, and two additional negative real roots $\xi_2(\epsilon) \leq  -1 \leq \xi_3(\epsilon) \leq 0$ for $\epsilon \geq 3/2$. The relevant limiting values are $\xi_1(\epsilon \to -\infty) = -\epsilon^{-1}$,  $\xi_1(0)=2^{1/3}$, and $\xi_2(3/2) =1$.


The minimal distance is
\begin{align}
    d_{\text{min}}/d_0 =
    \begin{cases} 
    \kappa^{-2/3}   \, \xi_1(- \kappa^{-2/3}  \, E_{+}/ U) 
    \, , & E_{+} < 3 U/2  \, ,\\
    -  \xi_2( E_{+}/ U ) 
      \, , & E_{+} > 3 U/2 \, .
    \end{cases} \label{eq:dmin}
\end{align}

For $\kappa \sim 1$ and for $\kappa \ll 1$, $d_{\text{min}} \gtrsim d_0$ for all $E_{+}$. In particular, in the narrow constriction limit, $\kappa \ll 1$,
\begin{align}
d_{\text{min}}/d_0 \approx \begin{cases} 
\frac{U }{E_{+}} \, ,  & U_{\text{1D}} < E_{+}  < 3U/2 \, ,  \\
\sim \kappa^{-2/3}   \, , &  - U_{\text{1D}} < E_{+} < U_{\text{1D}}  \, ,
\\
 \sqrt{-2 E_{+}/U} /\kappa   \, , &   E_{+} \ll  - U_{\text{1D}} \, , \\
 \gtrsim 1  \, , & 
E_{+} > 3 U/2  \, .
\end{cases}
\end{align}

In the wide constriction limit, $\kappa \gg 1$, there is a range of values of $E_{+}$ such that  $d_0 > d_{\text{min}} >  2^{1/3} \, \kappa^{-2/3} d_0 $
for $- \kappa^2 U /2 < E_{+} < 3 U/2$.

%

\end{document}